# Scattering of ultrashort electron wave packets: optical theorem, differential phase contrast and angular asymmetries

Yuya Morimoto[1,2] and Lars Bojer Madsen[3]

[1] *RIKEN Cluster for Pioneering Research (CPR) and RIKEN Center for Advanced Photonics (RAP), RIKEN, Wako 351-0198, Japan*

[2] *Department of Nuclear Engineering and Management, Graduate School of Engineering, The University of Tokyo, 7-3-1 Hongo, Bunkyo-ku, Tokyo 113-8656, Japan*

[3] *Department of Physics and Astronomy, Aarhus University, 8000 Aarhus C, Denmark*

*yuya.morimoto@riken.jp

*bojer@phys.au.dk

**Recent advances in electron microscopy allowed the generation of high-energy electron wave packets of ultrashort duration. Here we present a non-perturbative S-matrix theory for scattering of ultrashort electron wave packets by atomic targets. We apply the formalism to a case of elastic scattering and derive a generalized optical theorem for ultrashort wave-packet scattering. By numerical simulations with 1-fs wave packets, we find in angular distributions of electrons on a detector one-fold and anomalous two-fold azimuthal asymmetries. We discuss how the asymmetries relate to the coherence properties of the electron beam, and to the magnitude and phase of the scattering amplitude. The essential role of the phase of the exact scattering amplitude is revealed by comparison with results obtained using the first-Born approximation. Our work paves a way for controlling electron-matter interaction by the lateral and transversal coherence properties of pulsed electron beams.**

## I. Introduction

Collision of electron beams with matter form the foundation of electron microscopy, electron-beam nanofabrication, and radiotherapy. Decades of developments in electron microscopy with advances in aberration correction allowed the generation of electron beams focused down to sub-Angstrom spot size [1–3]. Scanning transmission electron microscopy with a small probe beam provided atomic-scale images of electric as well as magnetic fields [4–7]. In parallel with the advancement of the transversal beam control, the control over the lateral size, i.e., the temporal duration, has approached the atomic scale [8–16]. The generation of attosecond-short electron beams was achieved via the velocity modulation of electron beams by light waves [8–16]. Pulses as short as 0.27 fs at the kinetic energy of 28 keV, corresponding to 27 nm lateral size, has been obtained [12], and efforts to produce shorter pulses are currently being pursued. The combined attosecond-Angstrom electron beam is within experimental reach [17].

These technical advances trigger the question whether there is any difference in scattering of ultrafast electron wave packets from those of long-pulsed or continuous beams. It was predicted that the excitation probabilities by an electron beam from a target can be enhanced when the beam is properly shaped in time [18–22]. Quantum-state measurements with short electron beams were also proposed [23–25].

In this work, by referring seminal works [26–28], we discuss a fully three dimensional nonperturbative S-matrix scattering theory where an electron beam is considered as a wave packet. To elucidate the theory, we perform numerical simulations of elastic scattering processes by isolated atoms. Through the comparison with the results from the perturbative Born approximation, we show the importance of the nonperturbative treatment in the description of the conservation of flux and in the description of the angular asymmetry in the scattering.

The paper is organized as follows. In Sec. II, the general formulation as well as the case of potential scattering is presented. In Sec. III the results are discussed. Finally, Sec. IV concludes. The Appendix contains detailed information and derivations to support the main text. Atomic units are used throughout unless indicated otherwise.

## II. Formulation

In this section, we discuss the S-matrix theory for the scattering of ultrashort electron wave packets. We first present a general formula. Then, the formula is simplified to the case of potential scattering and an optical theorem generalized to the ultrashort wave packets is derived.

### II-1. General formula

We consider the collisional process,

$$e(\boldsymbol{k}_i) + A(\boldsymbol{k}_{A,i}, n_i) \to e(\boldsymbol{k}_f) + A(\boldsymbol{k}_{A,f}, n_f), \qquad (1)$$

where $\boldsymbol{k}_i$ ($\boldsymbol{k}_f$) is the initial (final) wave number of the projectile electron, and $\boldsymbol{k}_{A,i}$ ($\boldsymbol{k}_{A,f}$) is the initial (final) wave number of the target atom. We consider high-energy (> keV) projectile electrons and exchange scattering is neglected, which is accurate at the considered energy [29]. In Eq. (1), $n_i$ ($n_f$) describes the initial (final) internal electronic state of the target. The total Hamiltonian of this system reads

$$H = H_0 + H_1 + V, \qquad (2)$$

where $H_0$ is for the projectile electron, $H_1$ is for the target, and $V$ is their interaction. When the projectile electron is far from the target, $V \approx 0$. By following [27], we consider eigenstates of the total Hamiltonian with outgoing wave boundary condition,

$$\Psi_i^+ = \Psi_i + G^+(\varepsilon_i)V\Psi_i, \qquad (3)$$

where $\Psi_i$ is an eigenfunction of $H_0 + H_1$ (i.e., $V = 0$), $G^+(\varepsilon_i) = \lim_{\eta \to +0} 1/(\varepsilon_i + i\eta - H)$ is the full Green's function, and $\varepsilon_i$ is the energy given by

$$\varepsilon_i = \frac{k_i^2}{2} + \frac{k_{A,i}^2}{2M} + \omega_{n_i}, \qquad (4)$$

where $M$ is the rest mass of the target, and $\omega_{n_i}$ is the eigenenergy of the target's electronic state $\psi_{n_i}$.

We allow the target to be prepared in a superposition of eigenstates with amplitudes $c_{n_i}$ at time $t_0$. The wave function of the target at time $t_p$ ($p$ stands for probe) when the projectile collides with the target is described by

$$\psi(t_p) = \sum_{n_i} c_{n_i} e^{-i\omega_{n_i}(t_p - t_0)} \psi_{n_i}. \qquad (5)$$

This description was also used in [26] for the ultrafast pump (i.e., excitation)-probe scheme. We note that this description of the target is general and could include electronic as well as nuclear degrees of freedom of the target.

We allow both the projectile and target to be prepared with momentum wave functions $a_e(\mathbf{k}_i)$ and $a_A(\mathbf{k}_{A,i})$ at wavenumbers $\mathbf{k}_i$ and $\mathbf{k}_{A,i}$. These wave functions in $\mathbf{k}$-space account for the spatial localization of the projectile and target, in other words, the focusing or divergence of the beams. The real-space density of the projectile electron at $t = t_p$ and position $\mathbf{r}$ is expressed as

$$\rho_e(\mathbf{r}) = \left| \frac{1}{\sqrt{(2\pi)^3}} \int d\mathbf{k}_i\, a_e(\mathbf{k}_i) e^{i\mathbf{k}_i \cdot \mathbf{r}} \right|^2. \qquad (6)$$

The total state reads

$$\Psi_{i,coh}(t_p) = \sum_{n_i} c_{n_i} e^{-i\omega_{n_i}(t_p - t_0)} \int d\mathbf{k}_i \int d\mathbf{k}_{A,i}\, a_e(\mathbf{k}_i) a_A(\mathbf{k}_{A,i}) \Psi_i^+. \qquad (7)$$

where the subscript coh indicates that the state is obtained by adding coherently all the wavenumbers contributing to the scattering state.

By generalization of the results from Ref. [27], the probability amplitude of the scattering into a final state with electron momentum $\mathbf{k}_f$ and target specified by $(\mathbf{k}_{A,f}, n_f)$ then reads

$$A_{\mathbf{k}_f, \mathbf{k}_{A,f}, n_f}(t_p) = \sum_{n_i} c_{n_i} e^{-i\omega_{n_i}(t_p - t_0)} \int d\mathbf{k}_i \int d\mathbf{k}_{A,i}\, a_e(\mathbf{k}_i) a_A(\mathbf{k}_{A,i})$$
$$\times \Bigg[ \delta(\mathbf{k}_f - \mathbf{k}_i)\delta(\mathbf{k}_{A,f} - \mathbf{k}_{A,i})$$
$$- 2\pi i \delta\left( \frac{k_f^2}{2} + \frac{k_{A,f}^2}{2M} + \omega_{n_f} - \frac{k_i^2}{2} - \frac{k_{A,i}^2}{2M} - \omega_{n_i} \right) \delta(\mathbf{k}_f + \mathbf{k}_{A,f} - \mathbf{k}_i$$
$$- \mathbf{k}_{A,i}) T_{\mathbf{k}_f, \mathbf{k}_{A,f}, n_f, \mathbf{k}_i, \mathbf{k}_{A,i}, n_i} \Bigg], \tag{8}$$

where $T_{\mathbf{k}_f, \mathbf{k}_{A,f}, n_f, \mathbf{k}_i, \mathbf{k}_{A,i}, n_i}$ denotes the exact T-matrix element for the scattering between the states specified by the subscripts. If the target is placed at the impact parameter of $\mathbf{b}$ [28], the result in Eq. (8) generalizes to

$$A_{\mathbf{k}_f, \mathbf{k}_{A,f}, n_f}(t_p, \mathbf{b}) = \sum_{n_i} c_{n_i} e^{-i\omega_{n_i}(t_p - t_0)} \int d\mathbf{k}_i \int d\mathbf{k}_{A,i}\, a_e(\mathbf{k}_i) a_A(\mathbf{k}_{A,i}) e^{-i\mathbf{k}_{A,i}\cdot\mathbf{b}}$$
$$\times \Bigg[ \delta(\mathbf{k}_f - \mathbf{k}_i)\delta(\mathbf{k}_{A,f} - \mathbf{k}_{A,i})$$
$$- 2\pi i \delta\left( \frac{k_f^2}{2} + \frac{k_{A,f}^2}{2M} + \omega_{n_f} - \frac{k_i^2}{2} - \frac{k_{A,i}^2}{2M} - \omega_{n_i} \right) \delta(\mathbf{k}_f + \mathbf{k}_{A,f} - \mathbf{k}_i$$
$$- \mathbf{k}_{A,i}) T_{\mathbf{k}_f, \mathbf{k}_{A,f}, n_f, \mathbf{k}_i, \mathbf{k}_{A,i}, n_i} \Bigg]. \tag{9}$$

The integral over $\mathbf{k}_{A,i}$ can be evaluated by using the properties of the Dirac delta functions in Eq. (9), and the amplitude $A_{\mathbf{k}_f, \mathbf{k}_{A,f}, n_f}(t_p, \mathbf{b})$ is then expressed as

$$A_{\mathbf{k}_f, \mathbf{k}_{A,f}, n_f}(t_p, \mathbf{b}) = \sum_{n_i} c_{n_i} e^{-i\omega_{n_i}(t_p - t_0)} \int d\mathbf{k}_i\, a_e(\mathbf{k}_i)$$
$$\times \Bigg[ \delta(\mathbf{k}_f - \mathbf{k}_i) a_A(\mathbf{k}_{A,f}) e^{-i\mathbf{k}_{A,f}\cdot\mathbf{b}}$$
$$- 2\pi i a_A(\mathbf{k}_f + \mathbf{k}_{A,f} - \mathbf{k}_i) e^{-i(\mathbf{k}_f + \mathbf{k}_{A,f} - \mathbf{k}_i)\cdot\mathbf{b}} \delta\left( \frac{k_f^2}{2} + \frac{k_{A,f}^2}{2M} + \omega_{n_f} - \frac{k_i^2}{2} \right.$$
$$\left. - \frac{|\mathbf{k}_f + \mathbf{k}_{A,f} - \mathbf{k}_i|^2}{2M} - \omega_{n_i} \right) T_{\mathbf{k}_f, \mathbf{k}_{A,f}, n_f, \mathbf{k}_i, n_i} \Bigg]. \tag{10}$$

To calculate the probability for ending up at a final state, we consider $\left| A_{\mathbf{k}_f, \mathbf{k}_{A,f}, n_f}(t_p, \mathbf{b}) \right|^2$ and integrate or sum it over unobserved states.

## II-2. Potential scattering

To simplify the above equation, we consider a case where the widely-used approximation of potential scattering is applied, that is, a scenario is considered where the target is much heavier than the electron ($M \gg 1$) and the target is localized in space, corresponding to a broad momentum distribution much wider than that of electrons, $a_A(\boldsymbol{k}_f + \boldsymbol{k}_{A,f} - \boldsymbol{k}_i) \approx a_A(\boldsymbol{k}_{A,f})$ [26,27]. The non-perturbative scattering amplitude $f_{\boldsymbol{k}_f, n_f, \boldsymbol{k}_i, n_i}$ for the potential scattering of a plane-wave projectile is related to the T-matrix via

$$f_{\boldsymbol{k}_f, n_f, \boldsymbol{k}_i, n_i} = -(2\pi)^2 T_{\boldsymbol{k}_f, n_f, \boldsymbol{k}_i, n_i}. \tag{11}$$

The norm squared of the scattering amplitude relates directly to the scattering cross section. The coefficient between the scattering amplitude and the T-matrix element depends on the normalization of the continuum wave function. Our choice of momentum normalization with a factor of $1/(2\pi)^{3/2}$ on the plane waves leads to the factor of $(2\pi)^2$ in Eq. (11). Under these assumptions, the probability amplitude from Eq. (10) is expressed as

$$A_{\boldsymbol{k}_f, \boldsymbol{k}_{A,f}, n_f}(t_p, \boldsymbol{b})$$
$$= \sum_{n_i} c_{n_i} e^{-i\omega_{n_i}(t_p - t_0)} a_A(\boldsymbol{k}_{A,f}) e^{-i\boldsymbol{k}_{A,f} \cdot \boldsymbol{b}}$$
$$\times \left[ a_e(\boldsymbol{k}_f) + \frac{i}{2\pi} \int d\boldsymbol{k}_i \, a_e(\boldsymbol{k}_i) e^{-i(\boldsymbol{k}_f - \boldsymbol{k}_i) \cdot \boldsymbol{b}} \delta\left(\frac{k_f^2}{2} + \omega_{n_f} - \frac{k_i^2}{2} - \omega_{n_i}\right) f_{\boldsymbol{k}_f, n_f, \boldsymbol{k}_i, n_i} \right]. \tag{12}$$

Below we limit our discussion to elastic scattering by a target in the ground state ($n_f = n_1$, $c_{n_i} = \delta_{n_i,1}$). By using the relationship

$$\delta\left(\frac{k_f^2}{2} - \frac{k_i^2}{2}\right) = \frac{1}{k_f} \delta(k_f - k_i), \tag{13}$$

the probability amplitude for the elastic case then reads

$$A_{\boldsymbol{k}_f, \boldsymbol{k}_{A,f}, n_1}(t_p, \boldsymbol{b})$$
$$= e^{-i\omega_{n_1}(t_p - t_0)} a_A(\boldsymbol{k}_{A,f}) e^{-i\boldsymbol{k}_{A,f} \cdot \boldsymbol{b}} \left[ a_e(\boldsymbol{k}_f) + \frac{ik_f}{2\pi} \int d\hat{\boldsymbol{k}}_i \, a_e(k_f \hat{\boldsymbol{k}}_i) e^{-i(\boldsymbol{k}_f - k_f \hat{\boldsymbol{k}}_i) \cdot \boldsymbol{b}} f_{elas} \right] \tag{14}$$

where $f_{elas}$ is the elastic scattering amplitude for a plane-wave electron of $\boldsymbol{k}_i = k_f \hat{\boldsymbol{k}}_i \to \boldsymbol{k}_f$. The norm square of the probability amplitude is now independent of $t_p$. We assume that the final momentum of the target $\boldsymbol{k}_{A,f}$ is unresolved. By using the normalization condition $\int d\boldsymbol{k}_{A,f} \, |a_A(\boldsymbol{k}_{A,f})|^2 = 1$, the total probability for ending up at the final momentum $\boldsymbol{k}_f$ can be expressed as

$$\int d\boldsymbol{k}_{A,f} \left| A_{\boldsymbol{k}_f, \boldsymbol{k}_{A,f}, n_1}(\boldsymbol{b}) \right|^2 = I_{tot,\boldsymbol{b}}(\boldsymbol{k}_f) = I_{unsc}(\boldsymbol{k}_f) + I_{int,\boldsymbol{b}}(\boldsymbol{k}_f) + I_{sc,\boldsymbol{b}}(\boldsymbol{k}_f), \tag{15}$$

where

$$I_{unsc}(\boldsymbol{k}_f) = |a_e(\boldsymbol{k}_f)|^2, \tag{16}$$

$$I_{int,b}(\bm{k}_f) = -\frac{1}{\pi}\text{Im}\left[a_e^*(\bm{k}_f)k_f\int d\widehat{\bm{k}}_i\, a_e(k_f\widehat{\bm{k}}_i)e^{-i(\bm{k}_f-k_f\widehat{\bm{k}}_i)\cdot\bm{b}}f_{elas}\right], \tag{17}$$

and

$$I_{sc,b}(\bm{k}_f) = \frac{1}{4\pi^2}\left|k_f\int d\widehat{\bm{k}}_i\, a_e(k_f\widehat{\bm{k}}_i)e^{ik_f\widehat{\bm{k}}_i\cdot\bm{b}}f_{elas}\right|^2, \tag{18}$$

respectively. $I_{unsc}(\bm{k}_f)$ represents the intensity profile of the un-scattered part, $I_{int,b}(\bm{k}_f)$ the interference between the two terms inside the square-bracket in Eq. (14), and $I_{sc,b}(\bm{k}_f)$ the profile of the scattered part. We note that although these three contributions have different origins, they cannot a priori be distinguished experimentally. In general, $I_{unsc}(\bm{k}_f)$ has the narrowest angular profile followed by $I_{int,b}(\bm{k}_f)$ and $I_{sc,b}(\bm{k}_f)$. Accordingly, the relative contributions depend on the final angles and therefore the position on a detector; see Fig. 1(a).

Next, we consider the condition for the total-flux conservation and derive a generalized optical theorem. The integral of Eq. (15) over the final electron momentum $\bm{k}_f$ gives

$$-\frac{1}{\pi}\int d\bm{k}_f\, \text{Im}\left[a_e^*(\bm{k}_f)k_f\int d\widehat{\bm{k}}_i\, a_e(k_f\widehat{\bm{k}}_i)e^{-i(\bm{k}_f-k_f\widehat{\bm{k}}_i)\cdot\bm{b}}f_{elas}\right]$$

$$+\frac{1}{4\pi^2}\int d\bm{k}_f\left|k_f\int d\widehat{\bm{k}}_i\, a_e(k_f\widehat{\bm{k}}_i)e^{ik_f\widehat{\bm{k}}_i\cdot\bm{b}}f_{elas}\right|^2 = 0 \tag{19}$$

where the probability conserving relations $\iint d\bm{k}_f d\bm{k}_{A,f}\left|A_{\bm{k}_f,\bm{k}_{A,f},n_f}(\bm{b})\right|^2 = \int d\bm{k}_f\left|a_e(\bm{k}_f)\right|^2 = 1$ were used. This leads to the optical theorem generalized for a wave-packet projectile,

$$\int d\bm{k}_f\, k_f^2\left|f'_{elas}(\bm{k}_f,\bm{b})\right|^2 = 4\pi\,\text{Im}\left[\int d\bm{k}_f\, k_f a_e^*(\bm{k}_f)f'_{elas}(\bm{k}_f,\bm{b})\right], \tag{20}$$

where

$$f'_{elas}(\bm{k}_f,\bm{b}) = \int d\widehat{\bm{k}}_i\, a_e(k_f\widehat{\bm{k}}_i)e^{-i(\bm{k}_f-k_f\widehat{\bm{k}}_i)\cdot\bm{b}}f_{elas}, \tag{21}$$

which is the scattering amplitude averaged over incident angle $\widehat{\bm{k}}_i$ weighted by the momentum-space wave function $a_e(k_f\widehat{\bm{k}}_i)$.

Equations (20) and (21) describe an optical theorem under the situation where the incident electron beam has a broad momentum distribution such that different directions of the incoming momentum contribute to the same final momentum. The condition is more general than in the case of conventional plane-wave scattering, where interference between the scattered and un-scattered beams only occurs in the forward direction (i.e., zero scattering angle), and the conventional optical theorem reads

$$\sigma(k_0) = \frac{4\pi}{k_0}\text{Im}[f_{elas}(k_0,\theta_{0f}=0)]. \tag{22}$$

where $\bm{k}_0$ is chosen to be perpendicular to $\bm{b}$, $\theta_{0f}$ is the scattering angle in the electron frame,

and $\sigma(k_0)$ is the total elastic cross section at wave number $k_0$. In the Appendix A1, we show how the general expressions (20)-(21) simplify to Eq. (22) as they should for scattering with a monochromatic and well-collimated incoming wave.

### III. Result and discussion

Based on the formulas presented in Sec. II, we perform numerical simulations on the elastic scattering by isolated atoms. We discuss how the scattering depends on the atomic potential and the incoming electron beam properties.

### III-1. Simulation

Figure 1(a) illustrates the scattering geometry considered in this work. An electron wave packet propagating along $z$ is scattered by a target located at $\boldsymbol{b} = b_x \hat{\boldsymbol{x}}$ in the $z = 0$ plane. The location of the target is shifted with respect to the beam, which is equivalent to the situation of a fixed target and a scanned electron beam. Angular profiles of scattered and un-scattered electrons are detected by an imaging detector, and the final polar and azimuthal angles ($\theta_f$ and $\varphi_f$) are resolved. As in most of scanning transmission electron microscopy, the kinetic energy of the scattered electrons is not resolved [4,5,7]. Therefore, we consider an integral over the magnitude of the final momenta $k_f$ in the results to be presented below.

We consider a three-dimensional electron wave packet expressed in a simple way, namely an axially-symmetric wave packet whose momentum-space wave function is described by

$$a_e(\boldsymbol{k}_i) = \frac{1}{(8\pi^3)^{\frac{1}{4}} k_e \sqrt{\sigma_k} \sigma_\theta} \exp\left(-\frac{(k_i - k_e)^2}{4\sigma_k^2}\right) \exp\left(-\frac{\sin^2 \theta_i}{4\sigma_\theta^2}\right), \qquad (23)$$

where $\boldsymbol{k}_i$ is expressed in spherical coordinates ($k_i, \theta_i, \varphi_i$). This wave packet has gaussian distributions both in the total momentum $k_i$ and the incident angle $\theta_i$ with the rms widths of $\sigma_k$ and $\sigma_\theta$, respectively. Numerical integration gives that this wave function is normalized with the accuracy of $\int d\boldsymbol{k}_i |a_e(\boldsymbol{k}_i)|^2 = 0.9999$ with parameter values given below. We note that this wavefunction is analytically normalized, $\int d\boldsymbol{k}_i |a_e(\boldsymbol{k}_i)|^2 = 1$, when $k_i \approx k_i \cos \theta_i$ (limit of $\sigma_\theta \to 0$) and $k_i \approx k_e$ (limit of $\sigma_k \to 0$). The beam is focused both longitudinally and transversally in the plane of the target ($z = 0$). If needed, a chirp and higher-order dispersion of the wave packet can be included by multiplying an appropriate phase factor [30,31] on the expression in Eq. (23). Other angular profiles such as an aperture function [32,33] can also be adopted. In Eq. (23), $k_e$ is the length of the central wave number and we use in our simulations $k_e = 51$ Å$^{-1}$ corresponding to the kinetic energy of 10 keV and the speed of $v_e = 6 \times 10^7$ m/s. The width $\sigma_k$ is estimated from the relation to the inverse of the FWHM wave packet duration $\tau$ via $\tau = \sqrt{2\ln(2)}/(\sigma_k v_e)$ [34]. We consider a wave-packet duration of $\tau = 1$ fs, which is equivalent to the energy width of $\sigma_E = 0.8$ eV. $\sigma_\theta$ is the rms angular width and we use $\sigma_\theta = 10$ mrad (corresponding to 0.57 deg), a typical value for electron microscopes, unless otherwise specified. The corresponding rms spot size is approximately $1/(2k_e\sigma_\theta) = 1$ Å [34]. Figure 1(b) shows the image of un-scattered electrons, $\tilde{I}_{unsc}(\theta_f, \varphi_f) = \int dk_f\, k_f^2 I_{unsc}(\boldsymbol{k}_f)$ in a polar plot. The radius from the image center corresponds to the final polar angle $\theta_f$. The distribution in the figure is circularly symmetric around the center of the detector. The un-scattered profile is independent on the target and the impact parameter ($b_x$).

We consider three different atoms as targets, namely, hydrogen, helium, and argon. The target atoms are isolated and localized in space, corresponding to the broad momentum distribution over $\boldsymbol{k}_{A,i}$, see Sec. II-2. We employ two different elastic scattering amplitudes $f_{elas}$ for each target atom: amplitudes given by the nonperturbative Dirac partial-wave analysis [35,36] and the first-Born scattering amplitudes [37]. At the relatively high energy of 10 keV, the first-Born approximation provides a quantitatively accurate scattering cross sections, while it omits the phase of the amplitude (i.e., the imaginary part) originating from the phase induced by the atomic potential. In general, heavier targets provide larger phases of the scattering amplitude. Therefore, the comparison with the three target species and the two types of scattering amplitudes allows to elucidate the contribution of the scattering-amplitude phase and to discuss the validity of the first-Born approximation. At the relatively high energy of 10 keV, the scattering-amplitude phase is of the orders of 1 rad corresponding to a potential-induced delay of tens of zeptoseconds, see Appendix A2 for more details. Because of the small energy spread ($\sigma_k/k_e < 10^{-4}$), the change of $f_{elas}$ with respect to the variation of $k_i$ is very small (less than 0.01% variations of both the cross sections and phase per 1-eV energy shift) and thus neglected in the simulations, i.e., $f_{elas}$ is evaluated at $k_e$ only.

Figures 1(c) and (d) show simulated polar plots of the angular patterns, $\tilde{I}_{int,b_x}(\theta_f, \varphi_f) = \int dk_f\, k_f^2 I_{int,b_x}(\boldsymbol{k}_f)$ and $\tilde{I}_{sc,b_x}(\theta_f, \varphi_f) = \int dk_f\, k_f^2 I_{sc,b_x}(\boldsymbol{k}_f)$, respectively, for hydrogen as the atomic target. In these figures we employ scattering amplitudes given by the Dirac partial wave analysis. The interference profiles $\tilde{I}_{int,b_x}(\theta_f, \varphi_f)$ in Fig. 1(c) show a strong dependence on $b_x$. At $b_x = 0$, it is negative at all the final angles, satisfying the conservation of total flux (see below). At non-zero $b_x$, the interference profiles show vertical fringes, according to the factor $e^{-i(k_f - k_f \hat{k}_i) \cdot \boldsymbol{b}}$ in Eq. (17). Larger impact parameter (right panel) produces more fringes. These fringes cause an asymmetry in the total intensity with respect to $x = 0$ (i.e., right-left asymmetry in the figure) and are used for atom-by-atom imaging in the scanning transmission electron microscope [5]. Figure 1(d) show the patterns of scattered electrons $\tilde{I}_{sc,b_x}(\theta_f, \varphi_f)$ which are broader and nearly round. A small distortion (horizontal shrinkage) can be seen at the large impact parameter of $b_x = 3$ Å. At $\sigma_\theta = 10$ mrad and for the hydrogen atom target at $b_x =$ 1 Å, the contribution of the interference part $\tilde{I}_{int,b_x}(\theta_f, \varphi_f)$ is larger than that of the un-scattered part $\tilde{I}_{unsc}(\theta_f, \varphi_f)$ at $\theta_f > 2.7$ deg (47 mrad) and the scattered part $\tilde{I}_{sc,b_x}(\theta_f, \varphi_f)$ becomes the strongest above $\theta_f > 2.9$ deg (51 mrad). Below, we discuss details on the 2D profiles with particular attention to the flux conservation (i.e., optical theorem), the asymmetry of the interference profiles, and the angular asymmetry of the scattered electrons.

### III-2. Results on flux conservation

We evaluate the total probability of the scattered part $P_{sc}(b_x) = \int d\mathbf{k}_f\, I_{sc,b_x}(\mathbf{k}_f)$. As in experiments, we consider a detector with a limited angular range, which is $\theta_f$ up to 10 degrees (i.e., 175 mrad) in this work. In the limit of a plane-wave projectile, the detector covers 93% of the total scattering intensity from the hydrogen target. For the numerical evaluation of $P_{sc}(b_x)$, we first consider the integral over $\theta_f$ up to 10 degrees and then multiply the result with the factor 1/0.93. Red circles in Figs. 2(a) and 2(b) show total probabilities $P_{sc}(b_x)$ normalized to total incident flux as a function of the impact parameter. The signal resembles a gaussian with a rms width of 1.0 Å equivalent to the electron beam spot size. The results of the first-Born approximation illustrated by blue squares match well with those of the non-perturbative scattering amplitude, which demonstrates that the Born approximation provides accurate cross sections at the relatively high energy of 10 keV. On the other hand, the first-Born approximation violates the conservation of flux, as we now discuss. Red circles and blue squares in Figs. 2(c) and 2(d) show the total intensity of the interference term $P_{int}(b_x) = \int d\mathbf{k}_f\, I_{int,b_x}(\mathbf{k}_f)$. Nonperturbative simulations (red circles) give a gaussian-like distribution with a negative amplitude, but the results of the first-Born approximation are zero at all impact parameters. This difference can be understood by the fact that the elastic-scattering amplitudes given by the first-Born approximation are real-valued with vanishing imaginary part, whereas the exact scattering amplitude is complex and the phase is what gives the nonvanishing result in Figs. 2(c) and 2(d). Indeed, a non-zero imaginary value is required to satisfy the optical theorem and the conservation of the flux, see Eqs. (20)-(22). The magnitude of the nonperturbative results of $P_{int}(b_x)$ is nearly the same as that of $P_{sc}(b_x)$ in Fig. 2(a) at all $b_x$, illustrating the conservation of the total flux.

 To further verify the results of our simulation, we consider a model in which the target is represented by a circle on $z = 0$ whose radius $r_T = 0.03$ Å is given by the total scattering cross section ($2.9 \times 10^{-3}$ Å$^2$ [36]). In this model, electrons passing inside the circle are necessarily scattered while those passing outside are not. The total probability of the scattering part is thus estimated by $\int_{-\infty}^{+\infty}\int_{-\infty}^{+\infty} \Theta(x - b_x, y)\rho_e(x, y, z = 0)dxdy / \int_{-\infty}^{+\infty}\int_{-\infty}^{+\infty} \rho_e(x, y, z = 0)dxdy$, where $\Theta(x, y) = 1$ at $\sqrt{x^2 + y^2} < r_T$ and 0 elsewhere. The denominator is included for normalization and $\rho_e(x, y, z = 0)$ is the beam's profile on the plane of target, see Eq. (6). The results plotted by black curves in Figs. 2(a) and 2(b) well reproduce the simulation results. The black curves in Figs. 2(c) and 2(d) have the same profile but with opposite sign. The curves perfectly match with the simulations using nonperturbative amplitudes (red circles). This quantitative agreement certifies the accuracy of our simulations.

**III-3. Results on differential phase contrast**

Next, we discuss the differential phase contrast (DPC) which is used in high-resolution scanning transmission electron microscopy [5,32,33,38,39]. The DPC is often defined by the first moment or center of mass (CoM) of the total signal on a detector, which is expressed as

$$\text{CoM}x(b_x) = \int d\hat{\boldsymbol{k}}_f \, (\hat{\boldsymbol{x}} \cdot \hat{\boldsymbol{k}}_f) \, \tilde{I}_{tot,b_x}(\theta_f, \varphi_f)$$

$$\approx \int d\hat{\boldsymbol{k}}_f \, (\hat{\boldsymbol{x}} \cdot \hat{\boldsymbol{k}}_f) \, \tilde{I}_{int,b_x}(\theta_f, \varphi_f), \tag{24}$$

where we assume that the electron beam is scanned along the $x$ direction and the target is shifted by $b_x$, $d\hat{\boldsymbol{k}}_f = \sin\theta_f \, d\theta_f d\varphi_f$, $\hat{\boldsymbol{x}} \cdot \hat{\boldsymbol{k}}_f = \sin\theta_f \cos\varphi_f$ and $\tilde{I}_{tot,b_x}(\theta_f, \varphi_f) = \tilde{I}_{unsc}(\theta_f, \varphi_f) + \tilde{I}_{int,b_x}(\theta_f, \varphi_f) + \tilde{I}_{sc,b_x}(\theta_f, \varphi_f)$. The un-scattered part $\tilde{I}_{unsc}(\theta_f, \varphi_f)$ does not contribute to $\text{CoM}x(b_x)$ since it is circularly symmetric and independent of target. At small final polar angles, the contribution from the scattering part $\tilde{I}_{sc,b_x}(\theta_f, \varphi_f)$ is much weaker than the interference part $\tilde{I}_{int,b_x}(\theta_f, \varphi_f)$ for high-energy electrons, and thus $\tilde{I}_{int,b_x}(\theta_f, \varphi_f)$ is dominant; see Fig. 1. The angular profiles of the interference part at positive $b_x$ values shown in Fig. 1(b) are anti-symmetric around $x = 0$, which leads to the increase of total intensity at $x > 0$ and decrease at $x < 0$, or vice versa at negative values of $b_x$. Accordingly, the first moment of the total signal along $x$ becomes positive or negative. Classical mechanically, this effect is interpreted as the deflection of electron beams by atomic potentials [5,7]. High-energy electrons passing near an atom is attracted by the positively charged atomic core and their trajectories are deflected towards the side of the atom. As a result, the intensity profile on a detector is displaced toward the side of the atom, which agrees with the asymmetry seen above with quantum mechanical simulations. Figure 3(a) shows $\text{CoM}x(b_x)$ simulated with the nonperturbative amplitudes (red circles) and first-Born amplitudes (blue squares). Again, the target is hydrogen atom and $\sigma_\theta = 10$ mrad. The two simulations give nearly identical results. According to Eq. (17), the interference part is determined by the imaginary parts of $e^{-i(\boldsymbol{k}_f - k_f \hat{\boldsymbol{k}}_i) \cdot \boldsymbol{b}}$ and $f_{elas}$ given that we here assumed real-valued $a_e$. The accuracy of the first-Born approximation suggests that the phase $e^{-i(\boldsymbol{k}_f - k_f \hat{\boldsymbol{k}}_i) \cdot \boldsymbol{b}}$ originating from the impact parameter exerts the dominating contribution to DPC and $\text{CoM}x(b_x)$. The same conclusion is obtained with He and Ar targets (not shown) whose nonperturbative scattering amplitudes have larger imaginary values. In general, the wave function in momentum space $a_e(\boldsymbol{k}_i)$ is complex valued and when it has an angular (i.e., $\theta_i, \varphi_i$) dependent phase, the DPC is also influenced by the phase of $a_e(\boldsymbol{k}_i)$, see Eqs. (17) and (18).

We compare $CoMx(b_x)$ profiles with two different focusing angles, $\sigma_\theta = 10$ mrad [Fig. 3(a)] and $\sigma_\theta = 5$ mrad [Fig. 3(b)]. The profile with $\sigma_\theta = 10$ mrad is sharper than that of $\sigma_\theta = 5$ mrad since the beam spot size is smaller. The former profile is peaked at around $b_x = \pm 1$ Å while the latter is peaked at $b_x = \pm 2$ Å. It is known that when the beam spot size is larger than the size of atomic potential, the $CoMx(b_x)$ profile is approximately given by the spatial derivative of the beam profile $\partial \rho_e(x, y, z = 0)/\partial x$ [32], see Eq. (6) for the definition of $\rho_e(\mathbf{r})$. Black curves in Fig. 3 show such curves. They match well with the simulated $CoMx(b_x)$, showing that the atomic-resolution DPC with ultrashort electron beams can be explained in the same manner as that for continuous electron beams, and therefore the atomic-resolution ultrafast scanning electron microscopy is feasible when the spot size is smaller than the atomic separation.

### III-4. Results on angular asymmetry

The DPC analysis used above hides fine signal shapes on a detector. Therefore, we here discuss details of the angular profiles and their physical origins. First, to evaluate angular asymmetry of the profiles on a detector, we define an angle-resolved azimuthal contrast by

$$C_{tot,b_x}(\theta_f, \varphi_f) = \frac{\tilde{I}_{tot,b_x}(\theta_f, \varphi_f)}{\int_0^{2\pi} d\varphi_f \, \tilde{I}_{tot,b_x}(\theta_f, \varphi_f)/2\pi} - 1, \quad (25)$$

at a given impact parameter $b_x$, where the definition of the angular profile of the total signal $\tilde{I}_{tot,b_x}(\theta_f, \varphi_f)$ is given in Section III-1. The angle-resolved contrast $C_{tot,b_x}(\theta_f, \varphi_f)$ vanishes when the signal at a given $\varphi_f$ equals the signal averaged over all $\varphi_f$, a positive (negative) value means that the signal is larger (smaller) than the average. Figures 4(a)-(c) show results for $C_{tot,b_x}(\theta_f, \varphi_f)$ for the hydrogen atom at $b_x = 1, 2, 3$ Å, respectively. Note that there are no asymmetries at $b_x = 0$ [see Figs. 1(b)-(d)]. The nonperturbative scattering amplitudes are employed. Strong contrasts are found at around $\theta_f = 3$ deg (52 mrad) as ring-shaped patterns. With the increase of $b_x$, more complex structures appear. At $b_x = 1$ Å [Fig. 4(a)], there is a vertical node only at $x = 0$ (center of the panel) but at least three and five vertical nodes are seen at $b_x = 2$ Å [Fig. 4(b)] $b_x = 3$ Å [Fig. 4(c)], respectively.

The origins of the ring shapes and the vertical nodes are elucidated by considering the relative contributions of the three terms $\tilde{I}_{unsc}(\theta_f, \varphi_f)$, $\tilde{I}_{int,b_x}(\theta_f, \varphi_f)$, and $\tilde{I}_{sc,b_x}(\theta_f, \varphi_f)$ to the total intensity [Eq. (15)]. At small final polar angles, the un-scattered part $\tilde{I}_{unsc}(\theta_f, \varphi_f)$ is the strongest but it does not contribute to the asymmetry in the contrast because of its symmetric shape, see Fig. 1(b). However, it contributes to the denominator of Eq. (25). Because of its narrow angular profile [Fig. 1(b)], the denominator of Eq. (25) decreases with the final polar angles, and accordingly, the maximal contrast shows up at relatively large angles ($\theta_f = 3$ degrees) and draws an arc. The strongest contribution to the numerator is given by the interference part $\tilde{I}_{int,b_x}(\theta_f, \varphi_f)$ at small polar angles. The vertical nodes in the contrasts match well with those appearing in the angular profiles [Fig. 1(c)]. Again, the vertical fringes in Fig. 1(c) are attributed to the phase $e^{-i(\mathbf{k}_f - k_f \hat{\mathbf{k}}_i) \cdot \mathbf{b}}$ associated with the impact parameter, see Eq. (17). The fringe patterns are hidden in the center of mass analysis but clearly appears in the analysis of contrast here.

At larger final polar angles, the azimuthal contrast $C_{tot,b_x}(\theta_f, \varphi_f)$ shows different asymmetric patterns. Figures 4(d)-(f) plot the same results as in Figs. 4(a)-(c) but with magnified color scales to investigate relatively weak contrasts at larger polar angles ($\theta_f >$ 3 deg). Contrasts at these larger $\theta_f$ also show strong dependence on the impact parameter but weak dependence on $\theta_f$. At $b_x$ = 1 Å [Fig. 4(d)], the contrast shows an almost one-fold (i.e., right-left) asymmetry. On the contrary, the large-angle contrast at $b_x$ = 3 Å [Fig. 4(f)] shows an almost two-fold asymmetry. The one-fold asymmetric contrast attains positive values on the right-hand side ($0 < \varphi_f < \frac{\pi}{2}$ and $\frac{3\pi}{2} < \varphi_f < 2\pi$) while it is negative at the left side of the figure ($\frac{\pi}{2} < \varphi_f < \frac{3\pi}{2}$).

For the two-fold asymmetric contrast, intensities at $\frac{\pi}{4} < \varphi_f < \frac{3\pi}{4}$ (down), and $\frac{5\pi}{4} < \varphi_f < \frac{7\pi}{4}$, (up) are stronger (red) than those at $\frac{7\pi}{4} < \varphi_f < 2\pi$, $0 < \varphi_f < \frac{\pi}{4}$ (right) and $\frac{3\pi}{4} < \varphi_f < \frac{5\pi}{4}$ (left). The large-angle azimuthal contrast at $b_x$ = 2 Å [Fig. 4(e)] is somewhere between those of $b_x$ = 1 Å and $b_x$ = 3 Å.

The dependence on target atoms is also investigated. Results for the He and Ar are shown in Figs. 4(g)-(i) and Figs. 4(j)-(l), respectively. When the azimuthal contrasts at a given $b_x$ value are compared for the three different targets, for example, Fig. 4(f), (i), and (l) for $b_x$ = 3 Å, nearly the same patterns are observed at the small polar angles. This similarity is because the small-angle contrasts are mainly originating from the phase $e^{-i(\mathbf{k}_f - k_f \hat{\mathbf{k}}_i) \cdot \mathbf{b}}$ in $\tilde{I}_{int,b_x}(\theta_f, \varphi_f)$ which is common for all the targets. The quantitative differences are originating from $f_{elas}$ which depends on the target. On the other hand, the large-angle contrasts are strongly influenced by the choice of target. At $b_x$ = 3 Å, the two-fold contrast appears for the H atom [Fig. 4(f)] while the contrast for the Ar atom [Fig. 4(l)] is one-fold. Considering that the contribution of the scattered part $\tilde{I}_{sc,b_x}(\theta_f, \varphi_f)$ is strongest at the large polar angles (see Fig. 1), the target-dependent azimuthal contrasts should have origins in $\tilde{I}_{sc,b_x}(\theta_f, \varphi_f)$.

We therefore investigate the asymmetry of the scattering part alone by introducing the azimuthal contrast defined by

$$C_{sc,b_x}(\theta_f, \varphi_f) = \frac{\tilde{I}_{sc,b_x}(\theta_f, \varphi_f)}{\int_0^{2\pi} d\varphi_f \, \tilde{I}_{sc,b_x}(\theta_f, \varphi_f) / 2\pi} - 1, \qquad (26)$$

where the definition of the angular profile of the scattered part $\tilde{I}_{sc,b_x}(\theta_f, \varphi_f)$ is given in section III-1. To examine the importance of the scattering-amplitude phase, we compare results of the nonperturbative simulation with those by the Born approximation. Figure 5 shows $C_{sc,b_x}(\theta_f, \varphi_f)$ simulated for H, He, and Ar atoms with the nonperturbative amplitudes [panels in (a), (c), and (e)] and with first Born approximation [panels (b), (d), and (f)]. The results with the nonperturbative simulations in (a), (c), and (e) show profiles which can be seen as the result of superimposed signals of the one-fold (i.e., right-left) and the two-fold patterns, which is consistent with the results in Fig. 4 for the total signals. In contrast, as shown in panels (b), (d), and (f), the Born approximation always provides the two-fold patterns regardless of impact parameter and target atom, suggesting that the one-fold asymmetry in the contrast is associated with the phase of the scattering amplitudes while the two-fold asymmetry is not. As discussed above, the one-fold asymmetry in the contrast can be related to the classical interpretation of beam deflection by the atomic potential, but the two-fold asymmetry cannot. Although there exist reports indicating the two-fold asymmetry in the contrast [40,41], we know of no discussion of its quantum mechanical origin. Below, we show that both the one-fold and two-fold asymmetries in the contrasts of the scattered part originate from the spatial coherence of the electron beam and that the one-fold asymmetry is attributed to the phase of the scattering amplitude captured by a beyond first-Born treatment of the interaction of the projectile electron with the atomic potential.

When the results with nonperturbative amplitudes [Figs. 5 (a), (c), and (e)] are carefully compared with each other, we see a general trend: the one-fold asymmetry in the contrast is stronger than the two-fold asymmetry in the contrast at smaller impact parameter $b_x$ and with heavier target. For the H atom, the two-fold asymmetry is visible at $b_x = 1$ Å and dominant at $b_x = 3$ Å. On the other hand, for the Ar atom, the one-fold asymmetry in the contrast is dominant even at $b_x = 3$ Å. To our best knowledge, the two-fold asymmetry in the contrast was not observed experimentally. It is probably because samples containing relatively heavy atoms are usually chosen for electron microscopy. To quantify the two types of asymmetries in the contrasts, we define one-fold and two-fold azimuthal asymmetries as follows

$$A_1(b_x) = \int \sin\theta_f \, d\theta_f \frac{\int_0^{\pi/2} d\varphi_f \, \tilde{I}_{sc} + \int_{3\pi/2}^{2\pi} d\varphi_f \, \tilde{I}_{sc} - \int_{\pi/2}^{3\pi/2} d\varphi_f \, \tilde{I}_{sc}}{\int_0^{2\pi} d\varphi_f \, \tilde{I}_{sc}}, \quad (27)$$

$$A_2(b_x) = \int \sin\theta_f \, d\theta_f \frac{\int_0^{\pi/4} d\varphi_f \tilde{I}_{sc} + \int_{3\pi/4}^{5\pi/4} d\varphi_f \tilde{I}_{sc} + \int_{7\pi/4}^{2\pi} d\varphi_f \tilde{I}_{sc} - \int_{\frac{\pi}{4}}^{\frac{3\pi}{4}} d\varphi_f \tilde{I}_{sc} - \int_{5\pi/4}^{7\pi/4} d\varphi_f \tilde{I}_{sc}}{\int_0^{2\pi} d\varphi_f \tilde{I}_{sc}}, \quad (28)$$

where we suppress the argument of $\tilde{I}_{sc,b_x}(\theta_f, \varphi_f)$ for notational convenience. Equation (27) quantifies for each $\varphi_f$ the one-fold right-left asymmetry in Fig. 5 by a single number as a function of the $b_x$ value. Equation (28) does so for the two-fold (right-left, up-down) asymmetry. As in section III-2, we evaluate the integrals for $A_1(b_x)$ and $A_2(b_x)$ by considering $\theta_f$ up to 10 degrees (175 mrad) to reflect the experimental situation. Figure 6 shows results of the asymmetries $A_1(b_x)$ [Fig. 6(a)] and $A_2(b_x)$ [Fig. 6(b)] simulated for the three types of target atoms obtained with the two different methods (red circles are the nonperturbative scattering treatment and blue squares are the first-Born approximation). The magnitudes of the asymmetries range from a few percent to a few tens of percent. The one-fold asymmetry measure $A_1(b_x)$ is approximately linear in $b_x$ and the sign flips with the sign of $b_x$. On the other hand, the two-fold asymmetry parameter $A_2(b_x)$ is approximately quadratic in $b_x$ and the sign is always negative. As seen above, the relative magnitudes between the one-fold and the two-fold asymmetry depend on the target. Interestingly, the magnitudes of $A_2(b_x)$ are almost independent on the target species. The first-Born approximation gives a good estimate of $A_2(b_x)$ but not $A_1(b_x)$. For H, $|A_2(b_x)|$ is larger than $|A_1(b_x)|$ except for $b_x \sim 0$, while $|A_1(b_x)|$ is always larger than $|A_2(b_x)|$ for Ar.

In order to reveal the origins of the asymmetries and their linear and quadratic trends with respect to $b_x$, we derive analytical estimates for the asymmetries. Equation (18) for $I_{sc,b_x}(\mathbf{k}_f)$ contains integrals over $\hat{\mathbf{k}}_i$, or incident angles $\theta_i$ and $\varphi_i$, see Fig. 1(a). In addition, the two-dimensional image $\tilde{I}_{sc,b_x}(\theta_f, \varphi_f)$ in Fig. 1(d) is obtained via an integral over the length of final electron momentum $k_f$. However, we find by simulations that only one value of $k_i$ ($k_i = k_e$), one value of $\theta_i$ [$\theta_i = 0.8$ deg (14 mrad), that is the angle giving the strongest contribution to $I_{sc,b_x}(\mathbf{k}_f)$ in Eq. (18) due to $\sin\theta_i$ in $d\hat{\mathbf{k}}_i$ when $\sigma_\theta = 10$ mrad], and four values of $\varphi_i$ ($\varphi_i = 0, \frac{\pi}{2}, \pi$, and $\frac{3\pi}{2}$ where $\varphi_i$ is measured from $x$ axis) are sufficient to reproduce the azimuthal asymmetries in Fig. 5, see Appendix A3 and Fig. S2 for more details. If the number of $\varphi_i$ values is further reduced, it is not possible to reproduce the findings even qualitatively. Therefore, the scattering asymmetries originate from quantum interferences among multiple incident angular components of the incoming electron wave packet ending up in the same final angles; note that the magnitude square in Eq. (18) induces the coupling between two different components. Indeed, as we derive in Appendix A3, the difference in scattering intensities at $\varphi_i = 0$ and $\varphi_i = \pi$, which represents the degree of the one-fold asymmetry, with the four-trajectory assumption can be expressed as

$$\tilde{I}_{sc,b_x}(\theta_f, \varphi_f = 0) - \tilde{I}_{sc,b_x}(\theta_f, \varphi_f = \pi)$$

$$\propto \sum_{\varphi_i = \frac{\pi}{2}, \frac{3\pi}{2}} |f_{\varphi_i=0}||f_{\varphi_i}| \sin(\alpha_{\varphi_i=0} - \alpha_{\varphi_i}) \sin(k_e b_x \sin\theta_i)$$

$$+ \sum_{\varphi_i = \frac{\pi}{2}, \frac{3\pi}{2}} |f_{\varphi_i=\pi}||f_{\varphi_i}| \sin(\alpha_{\varphi_i} - \alpha_{\varphi_i=\pi}) \sin(k_e b_x \sin\theta_i)$$

$$+ |f_{\varphi_i=0}||f_{\varphi_i=\pi}| \sin(\alpha_{\varphi_i=0} - \alpha_{\varphi_i=\pi}) \sin(2 k_e b_x \sin\theta_i), \tag{29}$$

where the scattering amplitude $f_{\varphi_i} = |f_{\varphi_i}| e^{i\alpha_{\varphi_i}}$ is defined as that describing scattering from $(\theta_i, \varphi_i)$ to $(\theta_f, \varphi_f = 0)$ and where the equality $f_{\varphi_i=\frac{\pi}{2}} = f_{\varphi_i=\frac{3\pi}{2}}$ follows from the symmetry of the system. Equation (29) provides several important implications on the one-fold asymmetry. First, it consists of cross terms between different $f_{\varphi_i}$, suggesting that the asymmetry originates from the interference of different pathways. Second, the one-fold asymmetry appears only when the phase $\alpha_{\varphi_i}$ or imaginary part of the scattering amplitudes are considered, which answers why the first-Born approximation fails to reproduce the one-fold asymmetry. Third, these phases for the different scattering amplitudes that enter should not be equal. Therefore, the one-fold asymmetry is sensitive to the phase difference in scattering due to the difference in incident beam angles $\varphi_i$. Fourth, because of $\alpha_{\varphi_i=0} > \alpha_{\varphi_i=\frac{\pi}{2}} = \alpha_{\varphi_i=\frac{3\pi}{2}} > \alpha_{\varphi_i=\pi}$ (see Appendix A3), the sign of the one-fold asymmetry is determined by the sign of $b_x$ in the sine functions. Fifth, Taylor expansions of $\sin(k_e b_x \sin\theta_i)$ and $\sin(2k_e b_x \sin\theta_i)$ in Eq. (29) suggest that the degree of asymmetry is linearly proportional to $b_x$ at small $b_x$, which is consistent with the results in Fig. 6(a).

On the other hand, the difference in scattering intensities at $\varphi_i = 0$ and $\varphi_i = \pi/2$, which corresponds to a simple modelling of the two-fold asymmetry, is approximately expressed as

$$\tilde{I}_{sc,b_x}(\theta_f, \varphi_f = 0) - \tilde{I}_{sc,b_x}(\theta_f, \varphi_f = \frac{\pi}{2})$$

$$\propto \left\{ \left|f_{\varphi_i=\frac{\pi}{2}}\right| \left|f_{\varphi_i=\frac{3\pi}{2}}\right| - |f_{\varphi_i=0}||f_{\varphi_i=\pi}| \right\} \{1 - \cos(2 k_e b_x \sin\theta_i)\}. \tag{30}$$

As with the one-fold asymmetry, the equation consists of cross terms between different $\varphi_i$. Thus, the quantum interference is also essential for this asymmetry. In contrast to the one-fold asymmetry and Eq. (29), the two-fold asymmetry appears even when the phase of the scattering amplitude $\alpha_{\varphi_i}$ is neglected. This explains why the first-Born approximation succeeds to reproduce the asymmetry. We find numerically that the sign of $\left|f_{\varphi_i=\frac{\pi}{2}}\right|\left|f_{\varphi_i=\frac{3\pi}{2}}\right| - \left|f_{\varphi_i=0}\right|\left|f_{\varphi_i=\pi}\right|$ is always negative at all the conditions (except $\theta_i = 0$ and $\theta_f = 0$) and targets in this work. Therefore, the scattering intensities for $\varphi_f = \frac{\pi}{2}, \frac{3\pi}{2}$ are larger than those at $\varphi_f = 0, \pi$. The Taylor expansion of $1 - \cos(2k_e b_x \sin\theta_i)$ suggests that the two-fold asymmetry is quadratic in $b_x$ at small $b_x$, which is consistent with the results in Fig. 6(b). Although we assume a real-valued momentum-space wave function $a_e(\mathbf{k}_i)$ for the projectile electrons, this wave function is in general complex valued. When it has an angular dependent phase (for example depending on $\varphi_i$), the scattering intensity and the angular asymmetry can also be influenced by the phase, see Eqs. (18) and (A5) as well as the discussion in Appendix A3.

**IV. Summary and conclusion.**

In summary, we discussed a nonperturbative S-matrix theory for the scattering of ultrashort wave packets, which is applicable to both elastic and inelastic scattering and also to a target with intrinsic dynamics, i.e., time-dependent states. By applying the formalism to elastic scattering, we obtained a generalization of the optical theorem to the case of scattering with ultrashort electron pulses. We performed simulations on the elastic scattering of 1-fs-long high-energy three-dimensional wave packets by isolated atoms and found that the differential phase contrast with the ultrashort wave packet is explained well by the existing theory for continuous electron beams, suggesting the feasibility of the ultrafast scanning transmission electron microscopy. In the angular profiles of scattered electrons detected at large polar angles, we observed simultaneously one-fold and two-fold azimuthal asymmetries, both of which are attributed to the quantum interferences among multiple incident wave packet components that are detected at the same final angles. The interference occurs thanks to the transversal coherence of pulsed electron wave packets and could be used for estimating the degree of coherence or observing the spatial phase of a beam. The one-fold and two-fold asymmetries exhibit completely different dependences on the electron beam-atom distance (i.e., impact parameter) and the target atom species. Thus, the combined measurement of the two types of asymmetries might help precise determination of atomic positions at large distances or the observation of light atoms in electron microscopy. The comparison with the Born approximation highlighted the crucial role of the scattering phase induced by the atomic potential. Our simple model relates the one-fold asymmetry directly to the phase of the scattering amplitude, a quantity that is not accessible in conventional scattering experiments. The importance of the lateral scattering phase and the transversal coherence revealed in this work suggest future possibilities of controlling electron-matter collisional processes by the modulation of electron beams by electromagnetic fields.

# Appendix

## A-1. Derivation of Eq. (22).

Here we show how the optical theorem of Eq. (22) follows from the wave-packet treatment [Eqs. 20)-(21)] in the case corresponding to conventional potential scattering. To this end, we consider a plane-wave projectile corresponding to a delta function in momentum space

$$a_e(\boldsymbol{k_i}) = \delta(\boldsymbol{k_i} - \boldsymbol{k_0}) = \frac{\delta(k_i - k_0)\delta(\theta_i - \theta_0)\delta(\phi_i - \phi_0)}{k_i^2 \sin\theta_i} \quad (A1)$$

where $\boldsymbol{k_0}$ is chosen to be perpendicular to $\boldsymbol{b}$. Eq. (21) is expressed as

$$f'_{elas}(\boldsymbol{k_f}, \boldsymbol{b}) = \frac{f_{elas}(k_f, \theta_{0f}) e^{-i\boldsymbol{k_f}\cdot\boldsymbol{b}} \delta(k_f - k_0)}{k_f^2}, \quad (A2)$$

where $\theta_{0f}$ is scattering angle defined for the scattering from $\hat{\boldsymbol{k}}_0$ to $\hat{\boldsymbol{k}}_f$. The right-hand-side of Eq. (20) then reads

$$\text{Im}\left[\int d\boldsymbol{k_f}\, k_f a_e^*(\boldsymbol{k_f}) \frac{f_{elas}(k_f, \theta_{0f}) e^{-i\boldsymbol{k_f}\cdot\boldsymbol{b}} \delta(k_f - k_0)}{k_f^2}\right] = \text{Im}\left[\frac{f_{elas}(k_0, \theta_{0f} = 0)}{k_0}\right]. \quad (A3)$$

On the other hand, the left-hand-side of Eq. (20) reads

$$\frac{1}{4\pi}\int d\boldsymbol{k_f}\, k_f^2 \frac{|f_{elas}(k_f, \theta_{0f}) e^{-i\boldsymbol{k_f}\cdot\boldsymbol{b}} \delta(k_f - k_0)|^2}{k_f^4} = \frac{1}{4\pi}\int d\hat{\boldsymbol{k}}_f |f_{elas}(k_0, \theta_{0f})|^2 = \frac{1}{4\pi}\sigma(k_0), (A4)$$

where $\sigma(k_0)$ is the total elastic cross section at the momentum $k_0$. We then obtain the optical theorem for a plane-wave projectile, expressed by Eq. (22). This confirms the generality of the optical theorem for wave-packet scattering [Eqs. (20)-(21)].

## A-2. Phase of scattering amplitude and corresponding delay

Figure S1 compares the elastic scattering amplitudes $f = |f|e^{i\alpha}$ calculated with the Dirac partial-wave analysis (i.e., a nonperturbative method) [35,36] shown by red solid curves and the first-Born approximation [37] shown by blue dotted curves. At the relatively high kinetic energy of 10 keV, the first-Born approximation is accurate for the determination of the magnitudes of the scattering amplitudes $|f|$ [Fig. S1(a)]. However, the phase of the scattering amplitude $\alpha$ is always zero [Fig. S1(b)]. Considering the wavelength ($\lambda = 1.2 \times 10^{-11}$ m) and the speed ($v = 5.8 \times 10^7$ m/s) of the 10-keV electrons, a scattering amplitude phase of 1 rad corresponds to the time delay of 33 zeptoseconds induced by the atomic potential.

## A-3. Model for the one-fold and two-fold asymmetries

The simulation of the angular profile of the scattered part $\tilde{I}_{sc}(\theta_f, \varphi_f)$ requires integrals over $k_f$, $\theta_i$, and $\varphi_i$, see Eq. (18). We find numerically that even a single $k_f$ value ($k_f = k_e$), a single $\theta_i$ value [$\theta_i = 0.8$ deg, i.e., 14 mrad] and four values of $\varphi_i$ ($\varphi_i = 0, \frac{\pi}{2}, \pi, \frac{3\pi}{2}$) are enough to qualitatively reproduce the azimuthal contrast $C_{sc,b_x}(\theta_f, \varphi_f)$ [Eq. (26)]. In other words, the basic physics behind the angular asymmetries can be caught by considering four representative trajectories and their interferences. $\theta_i = 0.8$ degrees is chosen since this incident angle gives the largest contribution to $\tilde{I}_{sc}(\theta_f, \varphi_f)$. The results in Fig. S2 are for the comparison with Fig. 5(a). As in Fig. 5(a), the one-fold asymmetry in the contrast is dominant at $b_x = 1$ Å and the two-fold asymmetry in the contrast is dominant at $b_x = 3$ Å.

Based on this finding, we derive an analytical formula describing the one-fold and two-fold asymmetries. Equation (18) for one $\theta_i$ and four $\varphi_i$ values is expressed as

$$\tilde{I}_{sc,b_x}^{model}(\theta_f, \varphi_f) = I_0 \left| a_e(\theta_i, \varphi_i = 0) e^{-ik_\theta b_x} f_{elas,\varphi_i=0} + a_e(\theta_i, \varphi_i = \frac{\pi}{2}) f_{elas,\varphi_i=\frac{\pi}{2}} + a_e(\theta_i, \varphi_i = \pi) e^{ik_\theta b_x} f_{elas,\varphi_i=\pi} + a_e(\theta_i, \varphi_i = \frac{3\pi}{2}) f_{elas,\varphi_i=\frac{3\pi}{2}} \right|^2, \quad (A5)$$

where $I_0$ is a constant value, $k_\theta = k_e \sin \theta_i$ and $\boldsymbol{b}$ is along $x$-axis. Equation (A5) shows that the intensity profile in this simple model is determined by the magnitude squared of the four terms in the bracket and their interferences. Therefore, the angular profile of the scattered part is influenced by the relative phases of $a_e(\theta_i, \varphi_i)$ and $f_{elas,\varphi_i}$ of different $\varphi_i$ values, which suggests the possibility to control the intensity and asymmetry by the spatial phase modulation of electron beams.

Since $a_e$ of this work does not depend on $\varphi_i$ [Eq. (23)], Eq. (A5) can be simplified to

$$\tilde{I}_{sc,b_x}^{model}(\theta_f, \varphi_f) = I_C \left| e^{-ik_\theta b_x} f_{elas,\varphi_i=0} + f_{elas,\varphi_i=\frac{\pi}{2}} + e^{ik_\theta b_x} f_{elas,\varphi_i=\pi} + f_{elas,\varphi_i=\frac{3\pi}{2}} \right|^2 \quad (A6)$$

where $I_C = |a_e(\theta_i, \varphi_i = 0)|^2 I_0$ is a constant value. To evaluate the asymmetries, we calculate intensities to a fixed $\theta_f$ and $\varphi_f = 0, \pi/2, \pi, 3\pi/2$. Thanks to the symmetry of the system, each $f_{elas}$ in Eq. (A6) can be expressed by one of three scattering amplitudes, $f_L = |f_L|e^{i\alpha_L}$, $f_M = |f_M|e^{i\alpha_M}$, or $f_S = |f_S|e^{i\alpha_S}$ where $|f_L| \geq |f_M| \geq |f_S|$ and $\alpha_L \leq \alpha_M \leq \alpha_S$ depending on the scattering angles in the electron's frame. The assignment is summarized in Table I. For example, for $\varphi_f = 0$, $\varphi_i = \pi$ (i.e., close to straight trajectory) gives the smallest scattering angle in electron's frame and thus $f_L$ is chosen. On the other hand, the trajectory with $\varphi_i = 0$ is associated with large electron-frame scattering angle and $f_S$ is adopted. With these expressions, Eq. (A6) to final azimuthal angles $\varphi_f$ can be re-written as

$$\tilde{I}_{sc,b_x}^{model}(\theta_f, \varphi_f = 0) = I_C \left| e^{-ik_\theta b_x} f_S + 2f_M + e^{ik_\theta b_x} f_L \right|^2, \quad (A7)$$

$$\tilde{I}_{sc,b_x}^{model}(\theta_f, \varphi_f = \frac{\pi}{2}) = \tilde{I}_{sc,b_x}^{model}(\theta_f, \varphi_f = \frac{3\pi}{2}) = I_C |f_L + f_S + 2f_M \cos k_\theta b_x|^2, \quad (A8)$$

and

$$\tilde{I}_{sc,b_x}^{model}(\theta_f, \varphi_f = \pi) = I_C \left| e^{-ik_\theta b_x} f_L + 2f_M + e^{ik_\theta b_x} f_S \right|^2. \quad (A9)$$

Equation (A8) shows that there is no one-fold asymmetry perpendicular to the direction of target $\boldsymbol{b}$, i.e., between $\varphi_f = \frac{\pi}{2}$ and $\varphi_f = \frac{3\pi}{2}$. However, the one-fold asymmetry along $\boldsymbol{b}$ exits and the difference between $\tilde{I}_{sc,b_x}^{model}(\theta_f, \varphi_f = 0)$ and $\tilde{I}_{sc,b_x}^{model}(\theta_f, \varphi_f = \pi)$ is obtained as

$$\frac{\tilde{I}_{sc,b_x}^{model}(\theta_f, \varphi_f = 0) - \tilde{I}_{sc,b_x}^{model}(\theta_f, \varphi_f = \pi)}{4I_C}$$
$$= 2|f_S||f_M|\sin(\alpha_S - \alpha_M)\sin k_\theta b_x + 2|f_L||f_M|\sin(\alpha_M - \alpha_L)\sin k_\theta b_x$$
$$+ |f_S||f_L|\sin(\alpha_S - \alpha_L)\sin 2k_\theta b_x, \qquad (A10)$$

which is Eq. (29) in the main text.

Similarly, the expression for the two-fold asymmetry is obtained as

$$\frac{\tilde{I}_{sc,b_x}^{model}(\theta_f, \varphi_f = 0) - \tilde{I}_{sc,b_x}^{model}(\theta_f, \varphi_f = \frac{\pi}{2})}{2I_C}$$
$$= |f_M|^2\{1 - \cos(2k_\theta b_x)\} + 2|f_S||f_M|\sin(\alpha_S - \alpha_M)\sin k_\theta b_x$$
$$+ 2|f_L||f_M|\sin(\alpha_M - \alpha_L)\sin k_\theta b_x$$
$$- 2|f_S||f_L|\sin(k_\theta b_x - \alpha_S + \alpha_L)\sin k_\theta b_x. \qquad (A11)$$

Under the condition of $\alpha_L \approx \alpha_M \approx \alpha_S$, Eq. (A11) simplifies to

$$\frac{\tilde{I}_{sc,b_x}^{model}(\theta_f, \varphi_f = 0) - \tilde{I}_{sc,b_x}^{model}(\theta_f, \varphi_f = \frac{\pi}{2})}{2I_C} \approx \{|f_M|^2 - |f_S||f_L|\}\{1 - \cos(2k_\theta b_x)\}, \qquad (A12)$$

which is Eq. (30) in the main text.

Table I. Assignment of scattering amplitudes. Due to the symmetry of the system, the scattering amplitudes for the following combinations of $(\varphi_i, \varphi_f)$ at fixed $\theta_i$ and $\theta_f$ are one of the three amplitudes $f_L$, $f_M$, and $f_S$, see text for details.

|  | $\varphi_i = 0$ | $\varphi_i = \pi/2$ | $\varphi_i = \pi$ | $\varphi_i = 3\pi/2$ |
|---|---|---|---|---|
| $\varphi_f = 0$ | $f_S$ | $f_M$ | $f_L$ | $f_M$ |
| $\varphi_f = \pi/2$ | $f_M$ | $f_S$ | $f_M$ | $f_L$ |
| $\varphi_f = \pi$ | $f_L$ | $f_M$ | $f_S$ | $f_M$ |
| $\varphi_f = 3\pi/2$ | $f_M$ | $f_L$ | $f_M$ | $f_S$ |

**Acknowledgement**

We thank Mads Brøndum Carlsen for useful comments and discussions. YM acknowledges Peter Hommelhoff for his general support. This research was supported by JST ACT-X JPMJAX21AO, JST FOREST, MEXT/JSPS KAKENHI JP21K21344, Gordon and Betty Moore Foundation, Kazato Research Foundation, Research Foundation for Opto-Science and Technology, Yamada Science Foundation, FY2023 JSPS Invitational Fellowship for Research in Japan, and Independent Research Fund Denmark (Grant No. 9040-00001B and Grant No. 1026-00040B).

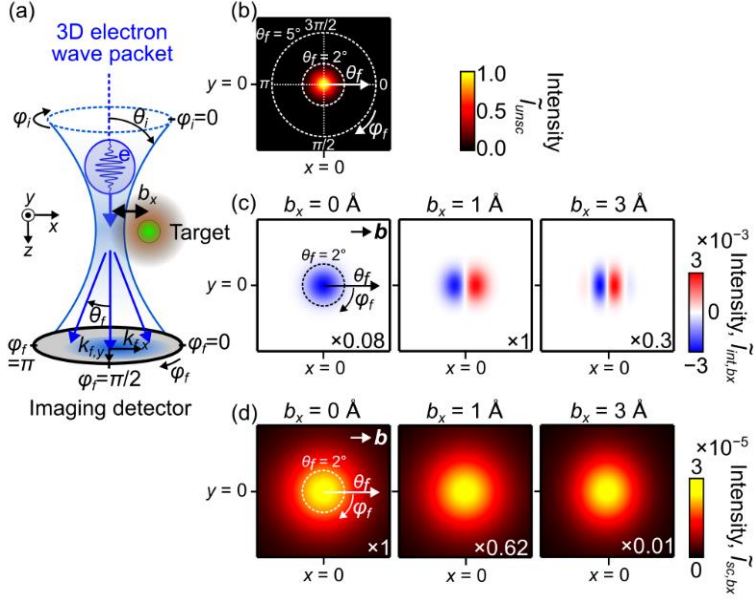

Fig.1. (a) Scattering geometry considered in this work. An electron wave packet (blue) is focused transversally and longitudinally at $z = 0$. A target (green and brown) placed at $x = b_x$, $y = z = 0$ scatters the wave packet. Scattered and un-scattered electrons are detected by an imaging detector. $\theta_i$ and $\varphi_i$ are incident polar and azimuthal angles, respectively. $\theta_f$ and $\varphi_f$ are final polar and azimuthal angles, respectively. These angles are defined in the laboratory frame. (b) Polar plot of the angular profile of the un-scattered part, see section III-1 for its definition. Peak intensity is normalized to 1. The radius corresponds to the final polar angle. White dotted circles show final polar angles of 2 degrees and 5 degrees (i.e., 35 mrad and 87 mrad). Final azimuthal angles of 0, $\pi/2$, $\pi$, and $3\pi/2$ are also shown. The center of the detector is at $x = y = 0$. (c) Polar plots of the angular profiles of the interference part for scattering on hydrogen, see section III-1 for its definition, at three different impact parameters. (d) Polar plots of the angular profiles of the elastically scattered part at the three different impact parameters for scattering on hydrogen, see section III-1 for its definition. In (c) and (d), the color scale is normalized in each plot and the scaling factor is expressed in the panel. The absolute intensity is normalized by the peak intensity of the un-scattered electrons in (b). See text for electron pulse parameters.

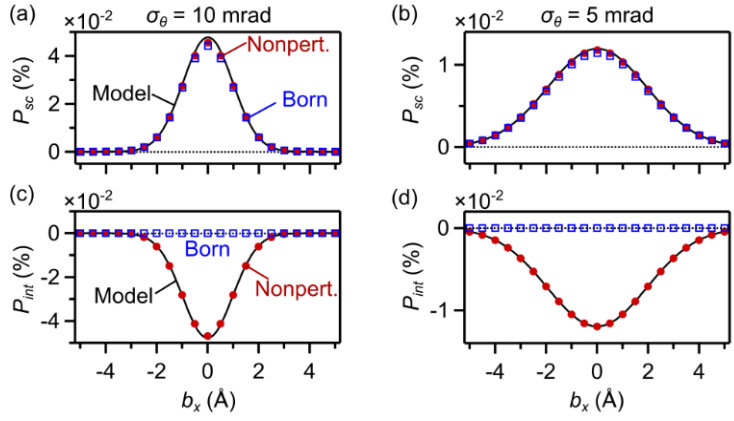

Fig. 2. Scattering probability and flux conservation for hydrogen as target. Plots in (a) and (b) show total elastic scattering probabilities $P_{sc}(b_x)$ (see section III-2 for its definition) normalized by the total incident flux at $\sigma_\theta = 10$ mrad (a) and $\sigma_\theta = 5$ mrad (b). Plots in (c) and (d) show total intensity of the interference term $P_{int}(b_x)$, see section III-2 for its definition. Red circles show results with nonperturbative scattering amplitudes. Blue open squares plot results with the first-Born approximation. Black solid curves show results of model calculations, see text for details. The total flux is conserved when the sum of (a) and (c) is zero; same for (b) and (d).

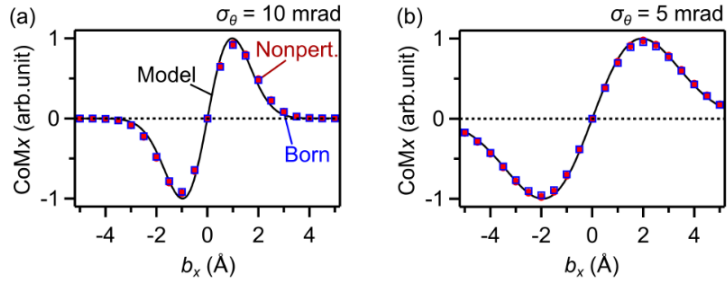

Fig. 3. Differential phase contrast and center of mass analysis for hydrogen as target. The center of mass $\text{CoMx}(b_x)$ is defined by Eq. (24) and calculated at $\sigma_\theta$ = 10 mrad (a) and $\sigma_\theta$ = 5 mrad (b). Red circles are results with nonperturbative calculations. Blue open squares are results with first Born approximation. Black solid curves show results of model calculations, see text for details. The scales of black curves are adjusted.

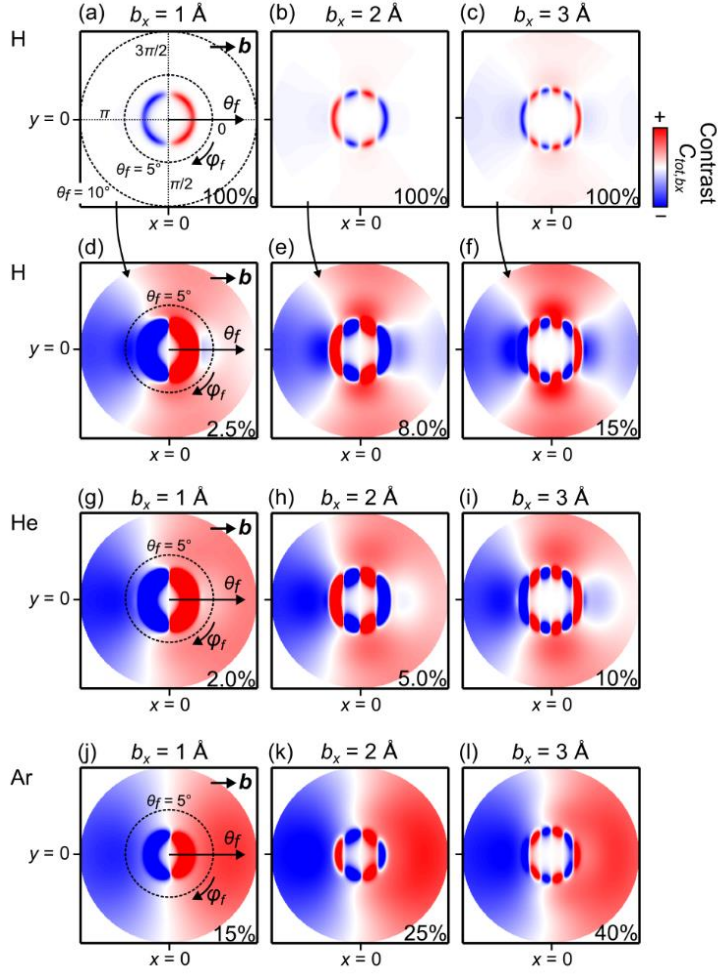

Fig. 4. Polar plots of the azimuthal contrast in total signals on a detector $C_{tot,b_x}(\theta_f, \varphi_f)$ defined by Eq. (25). (a)-(c) Azimuthal contrasts for the hydrogen target simulated with nonperturbative scattering amplitudes at $b_x$ = 1, 2, and 3 Å. Black dotted circles in (a) show final polar angles of 5 degrees (87 mrad) and 10 degrees (175 mrad). (d)-(f) Same results as in (a)-(c) but with different color scales. (g)-(i) Azimuthal contrasts for helium target. (j)-(l) Azimuthal contrasts for argon target. The limit of the color scale used in each plot is given as a percentage in each panel.

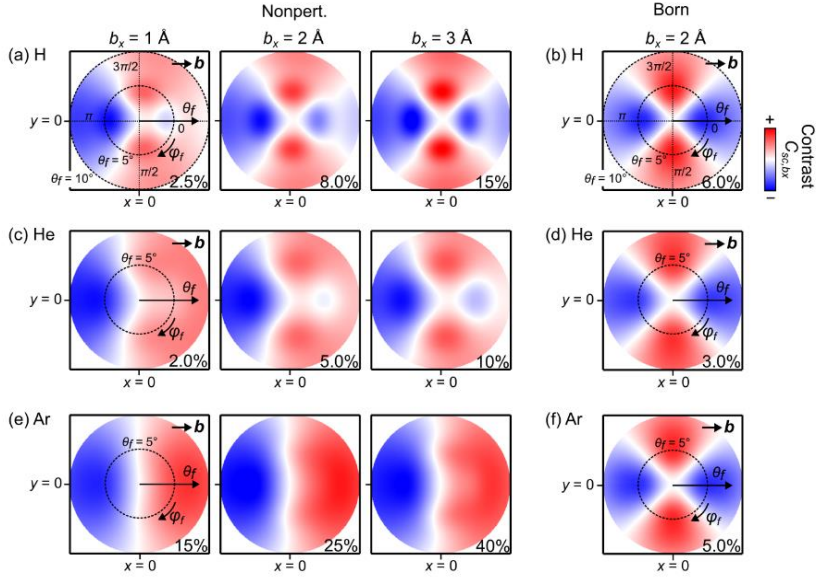

Fig. 5. Polar plots of the azimuthal contrast of the elastically scattered part alone, $C_{sc,b_x}(\theta_f, \varphi_f)$ defined by Eq. (26). (a), (c), (e) Angular contrasts of the scattered part simulated with nonperturbative scattering amplitudes at $b_x = 1, 2,$ and $3$ Å. The focusing angle is $\sigma_\theta = 10$ mrad. Target atoms are (a) hydrogen, (c) helium and (e) argon. The black dotted circles in (a) show the final polar angles of 5 degrees (87 mrad) and 10 degrees (175 mrad). The direction of the target displacement, i.e., the impact parameter vector, is indicted by black arrows. The color scale is normalized in each plot and the maximum contrast is given as a percentage in each panel. (b), (d), (f) Azimuthal contrasts of scattered part simulated with first Born approximation. The targets are (b) hydrogen, (d) helium and (f) argon. In (b), (d) and (f) only the 2-fold asymmetry in the contrast is observed. Since the angular profiles of the contrast are nearly identical at any non-zero $b_x$, only results at $b_x = 2$ Å are shown.

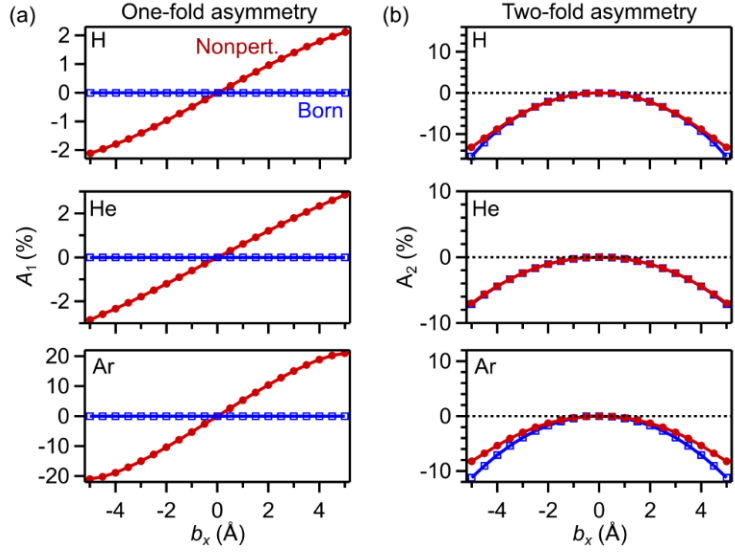

Fig. 6. Detailed analysis on the scattering asymmetry. (a) One-fold asymmetries $A_1(b_x)$ with respect to the impact parameter $b_x$. See Eq. (27) for the definition of the asymmetry. The target atoms are hydrogen (top), helium (middle) and argon (bottom). The focusing angle is $\sigma_\theta = 10$ mrad. Red filled circles show results with nonperturbative scattering amplitudes. Blue open squares show results with first-Born approximation. The one-fold asymmetry is approximately linear to $b_x$. The one-fold asymmetry appears when phase of scattering amplitudes is considered with the nonperturbative method. (b) Two-fold asymmetries $A_2(b_x)$ defined by Eq. (28). The two-fold asymmetry is approximately quadratic to $b_x$, and the first-Born approximation provides good estimates. Black dotted line are zero lines.

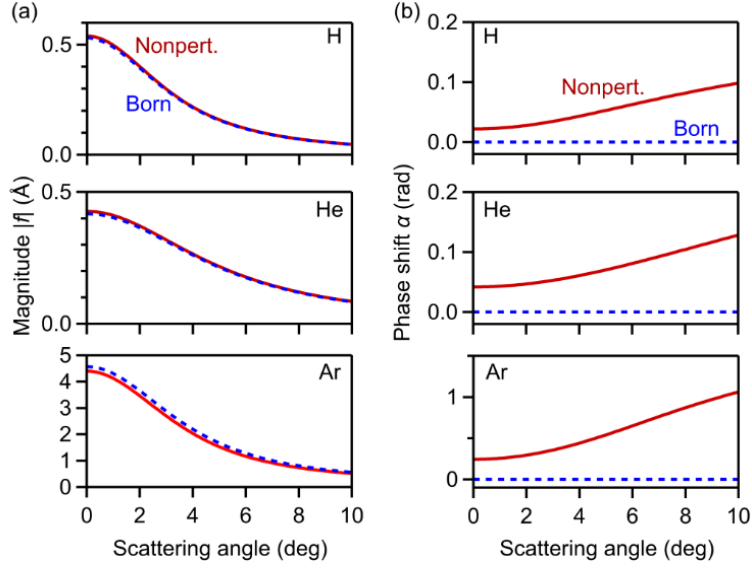

Fig. S1. Elastic scattering amplitudes. (a) Magnitudes of the elastic scattering amplitudes at 10 keV. Red solid curves show the results of nonperturbative calculations. Blue dotted curves show the results of the first-Born approximation. The scattering angle (horizontal axis) is defined in the frame of the electron. (b) Phases of the scattering amplitudes. A scattering-amplitude phase of 1 rad corresponds to the temporal delay of 33 zs.

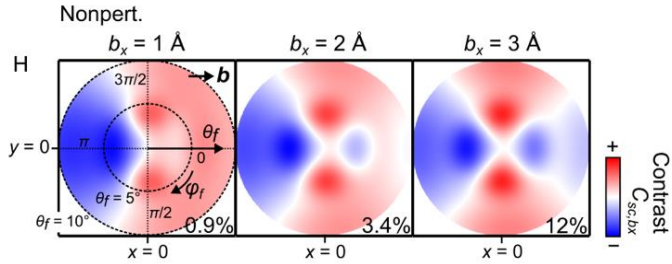

Fig. S2. Polar plots of the azimuthal contrast of the elastically scattered part $C_{sc,b_x}(\theta_f, \varphi_f)$ simulated with a reduced number of quantum pathways. The results are obtained by considering single $\theta_i$ value ($\theta_i = 0.8$ deg, 14 mrad) and four $\varphi_i$ values ($\varphi_i = 0, \pi/2, \pi,$ and $3\pi/2$). The other conditions are the same as those for Fig. 5(a) The color scale is normalized in each plot and the maximum asymmetries are given in percent.